\documentclass[preprint,aps,floats,showpacs,nofootinbib]{revtex4}
\usepackage{graphicx}
\usepackage{epsfig}
\usepackage{pstricks}
\usepackage{amsmath}
\usepackage{verbatim}

\def\be{\begin{equation}}
\def\ee{\end{equation}}
\def\bea{\begin{eqnarray}}
\def\eea{\end{eqnarray}}
{\newcommand{\lsim}{\mbox{\raisebox{-.6ex}{~$\stackrel{<}{\sim}$~}}}
{\newcommand{\gsim}{\mbox{\raisebox{-.6ex}{~$\stackrel{>}{\sim}$~}}}
\def\mpl{M_{\rm {Pl}}}

\def\gev{{\rm \,Ge\kern-0.125em V}}
\def\tev{{\rm \,Te\kern-0.125em V}}
\def\mev{{\rm \,Me\kern-0.125em V}}

\def\half{\frac{1}{2}}

\def\T1{T_1}

\def\phidot{{\dot{\phi}}}
\def\phiQ{\phi_Q}
\def\phistar{\phi_*}
\def\ph60{\phi_{*}}
\def\phie{\phi_e}
\def\calR={{\cal R}}
\def\calN={{\cal N}}

\begin{document}
\title{\bf 
The Quantum Phase of Inflation
}
\author{Arjun Berera} 
\email{ab@ph.ed.ac.uk}
\affiliation{
School of Physics and Astronomy, University of Edinburgh, Edinburgh EH9 3JZ, UK}
\author{Raghavan Rangarajan}
\email{raghavan@prl.res.in}
\affiliation{Theoretical Physics Division, Physical Research Laboratory,
Navrangpura, Ahmedabad 380 009, India}
\begin{abstract}
Inflation models can have an early phase of inflation where the evolution
of the inflaton is
driven by quantum fluctuations before entering the phase driven
by the slope of the scalar field potential. 
For a Coleman-Weinberg
potential this quantum phase lasts $10^{7-8}$ e-foldings.
A long period of fluctuation driven growth of the inflation field can possibly
take the inflaton 
past $\phi_{*}$, the value
of the field where our current horizon scale crosses the horizon;  
alternatively,
even if the field does not cross $\phi_{*}$, 
the inflaton could have high kinetic
energy at the end of this phase.  
Therefore
we study these issues in the context of 
different models of inflation.
In
scenarios where cosmological relevant scales leave during the
quantum phase 
we obtain large curvature perturbations of $O(10)$.
We also apply our results to quadratic curvaton models and to 
quintessence models.
In curvaton models we find that inflation must last longer than required to 
solve 
the horizon problem, that the curvaton models are incompatible with 
small field inflation
models and that there may be too large non-gaussianity.
A new phase of thermal fluctuation driven inflation is proposed,
in which during inflation the inflaton evolution is governed
by fluctuations from a sustained thermal radiation bath 
rather than by a scalar field potential.

\end{abstract}
\pacs{98.80.Cq}
\maketitle
\section{Introduction}
\noindent

In the early stages of inflation the evolution of the inflaton field
is dominated by
quantum fluctuations rather than by the slope of the potential, $V'(\phi)$.
In the initial stages of inflation,
$V'$ is small, and $\phidot$ is not given by
$-V'/(3H)$ but by the change in $\sqrt{\langle\phi^2\rangle}$ due to quantum fluctuations
leaving the horizon, freezing out, and becoming part of the large wavelength
background condensate \cite{BST,Linde}.
For a Coleman-Weinberg new inflation
potential
this initial phase can last for as long as
$10^{7-8}$ e-foldings after which $V'(\phi)$ dominates over
the fluctuations and one gets standard evolution.

In this article we discuss the quantum evolution of the inflaton
field in the context of new inflation, hilltop inflation, inflection point inflation,
chaotic inflation and natural inflation,
and then warm (new) inflation in the weak and strong dissipative regimes.
A long quantum
fluctuation driven phase for the inflaton can drive it to the bottom of 
its
potential or past $\phi_{*}$, the value of the inflaton field when our 
present horizon
scale crosses the horizon.  This will necessarily change our understanding
of inflation.  Alternatively, even if the inflaton does not cross $\phi_{*}$,
the inflaton kinetic energy at the end of the fluctuation driven phase can be 
large,
thereby precluding standard inflation thereafter.  
While the quantum phase of inflation has been known for long these 
possibilities have not been discussed in the literature so far.
The thermal fluctuation driven phase is new and for the first time is
being proposed in this paper.

We first consider
new inflation with a quartic potential, $V=V_0-\frac{1}{4}\lambda\phi^4$,
which mimics the inflationary part of a Coleman-Weinberg
potential when the field is far from the potential
minimum \cite{BST,Linde}.
We initially presume that our current horizon scale leaves the horizon
during the standard classical rolling phase (which fixes the coupling
$\lambda$ to be $\sim10^{-14}$)
and then check for the viability of this scenario after a long period
of quantum evolution.
We find that the field is still high up on its potential at the end
of the quantum phase, and its kinetic energy is not dominant.
Therefore the standard
new inflation scenario can follow an initial quantum fluctuation driven phase.

We then carry out the same analysis of quadratic new inflation, 
inflection point inflation, chaotic inflation and natural
inflation.
In chaotic, inflection point and natural inflation we find that, unlike
in new 
inflation, quantum fluctuations do not play a 
significant role
in the evolution of the inflaton, i.e., classical evolution effectively 
dominates from the 
beginning
of inflation.  

We next consider scenarios where our current horizon 
scale crosses the horizon during the quantum phase of inflation.  These are 
new scenarios that to our knowledge have not been considered earlier. 
We consider two scenarios - $\phi_*>\phi_Q>\phi_e$ and
$\phi_Q>\phi_e$, where 
$\phi_Q$ is the value of the inflaton
at the end of the quantum phase and
$\phi_e$ is the value of the inflaton
field at the end of inflation.
We do this for new inflation and hilltop inflation models.
The expressions for $\phi_{*}$ 
must be rederived for these scenarios in which
the current horizon scale leaves during the quantum phase. 
%
%
Since
$\dot\phi$ 
in 
is not given by
$\dot\phi = - V'/(3H)$ but by ${\sqrt{\langle\dot\phi^2\rangle}}$
the expression for the comoving curvature perturbation, ${\cal {R}}_k$, too must
be rederived.

For quartic new inflation models and $\phi_*>\phi_Q>\phi_e$ we find that 
consistency
of the scenario 
requires the coupling $\lambda$ to be
greater than $\sim 10^{-2}$.  
But this then implies that ${\cal {R}}_k\approx13$
for modes leaving during
the quantum phase.  Thus this scenario is ruled out.
This scenario for quadratic new inflation and hilltop inflation is also ruled 
out due
to conflict with the observed density perturbations.

For the scenario where the quantum
phase lasts the entire inflationary epoch, i.e. 
$\phi_Q>\phi_e$,
we find that it requires a very large value of $H\sim\mpl$
in new inflation models.
This scenario also generates too large curvature perturbations.

We then apply our results 
for modes leaving during
the quantum phase to quadratic curvaton models \cite{lythwands,LUW}, 
quintessence models 
and to a new model of dark energy \cite{ringevaletal}
where the dark energy is a condensate of quantum fluctuations generated during
inflation of a very light field.  
If quantum fluctuations of the curvaton determine the value of the curvaton field
at the end of
inflation then we find that the duration of inflation must be orders of magnitude
larger than the usual number of e-foldings required to solve the horizon 
problem.  
We also find that the slow roll parameter $\epsilon_H$ is determined in these 
models, 
and has a value larger than those compatible
with small field inflation models 
such as
new inflation, small field natural
inflation and some hybrid inflation
models
with a concave downward potential.
In the context of certain alternate expressions in the literature
for quantum fluctuations during quadratic chaotic inflation, 
the non-gaussianity in the curvaton models increases
to levels in conflict with observations.  
In the dark energy model we investigate whether the large curvature 
perturbations associated with the
dark energy condensate during inflation are compatible with the CMB constraints 
on the total
curvature perturbation
at decoupling.

We then consider the above issues in the warm inflationary scenario in
which the 
continuous 
decay of the inflaton during inflation creates a thermal bath which
survives throughout the inflationary phase \cite{Berera:1995ie}.
In warm inflationary dynamics, $\langle\phi^2\rangle$
can grow initially due to fluctuations of the inflaton field in this
thermal
background.  We study the quartic new inflation potential in the context
of both weak dissipation and strong dissipation of the inflaton field.
We investigate whether the inflaton reaches
the minimum of its potential
during the fluctuation driven phase itself.
We find that in the weak dissipative regime there is no thermal fluctuation
dominated phase.  Therefore the potential driven phase is preceded by a
quantum fluctuation driven phase, as in cold inflation.
In the strong dissipative regime there is a thermal fluctuation
driven phase which lasts for $10^8$ e-foldings.  This is a
new phase of warm inflation not considered earlier.
The further requirement that the thermal fluctuations do not take
the field to the bottom of the potential requires that the scale
of inflation is less than $10^{14}\gev$.  This is two orders of magnitude
less than earlier constraints \cite{BasteroGil:2009ec}.  
For both dissipative regimes,
after the fluctuation driven phase is over, 
standard warm inflation can follow,
indicating the consistency of the warm inflation scenario.

Below we summarize the paper.  In Sec. \ref{CI} we investigate the consistency of
cold inflation models preceded by a quantum phase.  In Secs. \ref{Qphase1} and 
\ref{Qphase2}
we consider scenarios in which the current horizon scale leaves the horizon 
during
the quantum phase of inflation.  In Sec. \ref{Otherscalars} we study the quantum 
phases of curvaton
and quintessence fields.  In Sec. \ref{WI} we apply our ideas to warm inflation 
and
investigate a thermal fluctuation driven phase.  Sec. \ref{Concl} contains our 
conclusions.

\section{Cold Inflation}\label{CI}

We first present the relevant equations for studying the quantum evolution phase 
in cold 
inflation models.  There are three scenarios that we consider - i) the
quantum phase ends before our current horizon scale crosses the 
horizon at time $t_*$, ii) the quantum phase ends after $t_*$ but before the end
of inflation at $t_e$, and iii) the quantum phase lasts for the entire duration
of inflation.  In the first scenario our focus is on determining whether the
inflaton ends the quantum phase with conditions suitable for a subsequent period
of classical inflation (thereby confirming the validity of existing models of 
inflation which allow a quantum phase).  For the other two cases, we 
investigate (in Secs. \ref{Qphase1} and \ref{Qphase2})
the feasibility of models where the current horizon scale leaves during
the quantum phase.



{\bf {$\phi_Q<\phi_*$}}

During the quantum evolution phase in cold inflation,
field fluctuations about an initial value
$\phi_0$ grow as
\be
\langle \delta\phi^2\rangle
=\frac{1}{(2\pi)^3}\int_H^{aH} {d^3k}   |\phi_k|^2
=\left(\frac{H}{2\pi}\right)^2 \times N(t) \ ,
\label{phisqcold}
\ee
where $k$ is the comoving momentum, $N(t)=H(t-t_0)$
is the number of e-foldings since the beginning of inflation at $t_0$,
and we have only integrated over modes outside the horizon which can act
as part of the homogeneous background field. 
For small field inflation models we 
ignore any initial $\phi_0$.
\footnote
{
For standard classical inflation one argues that due to quantum fluctuations the
initial value of the inflaton should not be less than $H/(2\pi)$.  This
argument is not relevant when one is studying quantum fluctuation driven
evolution. 
Nevertheless for small field models, presuming 
$\phi_0> H/(2\pi)$, 
we
impose
$\phi_*>H/(2\pi)$.
$\phi_0$ actually depends on the inflaton dynamics as the universe approaches
the inflationary epoch.  
}
(We have also assumed an effectively massless field, $m_{eff}\ll H$.)

Therefore
\be
\phi=\phi_0 \pm (H/2\pi)\sqrt{H(t-t_0)}
\label{phisqcoldwithphi0}
\ee
(we take $\phi_0>0$), and
$\phidot$ during the quantum phase is then given by
eq. (8.3.12) of Ref. \cite{Linde} as
\begin{equation}
\phidot_q = \frac{H^2}{4\pi \sqrt{H(t-t_0)}} \ .
\label{phdot}
\end{equation}

The period of quantum evolution lasts
while
\be
|\phidot_q|\gg|\phidot_V|=\left|-\frac{V'}{3H}\right|\,,
\label{quantumcondn}
\ee
that is, for values of $\phi$ satisfying
\be
\frac{3H^4}{8\pi^2}\gg |\phi-\phi_0)|\,| V'(\phi)|
\label{quantumcondn_a}
\ee
As we shall see, this condition is satisfied only for small field inflation
models such as new inflation.

To confirm that
the universe is potential energy
dominated during and at the end of the fluctuation driven epoch
we compare $\langle\dot\phi^2\rangle$
with $V(\phi)$.  Now
\be
\langle\phidot^2\rangle
=\frac{1}{(2\pi)^3}\int_H^{aH} {d^3k}   |\phidot_k|^2
\label{phidotsq}
\ee
Ignoring the gradient $(k^2/a^2)$ term in the equation of
motion for $\phi_k$,
\begin{equation}
\ddot\phi_k +3H\phidot_k + V^{\prime\prime}(\phi)\, \phi_k=0
\end{equation}
and,
\begin{equation}
\phidot_k \approx - \frac{V^{\prime\prime}(\phi)}{3H}\phi_k\,.
\end{equation}
Then
\begin{eqnarray}
\langle\phidot^2\rangle
&=&
\Biggr[\frac{V^{\prime\prime}(\phi)}{3H}\Biggr]^2
\frac{1}{(2\pi)^3}\int_H^{aH} {d^3k}   |\phi_k|^2\\
&=&
\Biggr[\frac{V^{\prime\prime}(\phi)}{3H}\Biggr]^2 \langle\delta\phi^2\rangle
\label{phidotsqnew}
\end{eqnarray}
We now apply the above results to specific models of inflation.

\subsection{Quartic new inflation}

For a Coleman-Weinberg potential,
modelled as 
$V=V_0 - \frac{\lambda}{4} \,\phi^4$ 
for small $\phi$, the quantum phase
occurs for $\phi<\phi_Q$, where, using eq. (\ref{quantumcondn}) and
eqs. (\ref{phisqcoldwithphi0}) and 
(\ref{phdot}),
\footnote{
Our value of $\phi_Q$ differs slightly from that obtained
in Ref. \cite{Linde}, possibly because of some factors in eq. (8.3.12)  of
Ref. \cite{Linde}.}
\be
\phi_Q= \frac{H}{2\pi} \left( \frac{60}{\lambda} \right)^{\frac1 4} 
\label{phQ}
\ee
for $\phi_0\approx0$.
Using eq. (\ref{phisqcoldwithphi0}), the number of e-foldings till $t_Q$ is 
$N_Q\equiv H(t_Q-t_0)=8/\sqrt{\lambda}$.
For the scenario with
$\phiQ<\phi_*$, $\lambda\sim 10^{-14}$ for GUT-scale inflation
and one can see that
$N_Q\approx10^8$, i.e., $10^8$ e-foldings
occur before one gets to classical evolution in this new inflation model.
$\phi_Q\approx 10^3H$. 
For $\phiQ<\phi_*$ the number of e-foldings of inflation
after the inflaton field crosses $\phi_*$ is
given by eq. (8.62) of Ref. \cite{KT}) as
\bea
{\cal N}(\phi_{*}\rightarrow\phi_e)
&=&\frac{8\pi}{\mpl^2} 
\int_{\phi_{*}}^{\phi_e} \frac{V}{-V'}{d\phi}
\label{N60}\\
&=&\frac{3H^2}{2\lambda}
\left(\frac{1}{\phi_*^2}-\frac{1}{\phi_e^2}\right)
\eea
and taking 
$\phi_e\approx(V_0/\lambda)^{1/4}$ we get
$\phi_*\approx10^6H$,
which is larger than $\phi_Q$ as presumed.
(Typically one uses the slow roll condition
$|V''|\ll9H^2$ to obtain $\phi_e=(3/\lambda)^{1/2} H$ \cite{KT}.  But then
$V(\phi_e)<0$.  
We instead take $\phi_e\approx (V_0/\lambda)^{1/4}$.)

$V(\phi_Q)=V_0-H^4/\pi^4$, and with $V_0\approx 0.1 H^2 \mpl^2$,
$V(\phiQ)\approx V_0$.
Using eq. (\ref{phidotsqnew}), $\langle\phidot^2\rangle=\lambda^2\phi^6/H^2$ 
and for $\phi\le\phi_Q$, $\langle\phidot^2\rangle\ll V_0$, i.e., conditions
for inflation remain valid during the quantum phase and at the end of the 
quantume phase.

The quantum evolution is statistical and thus
$\phi=(H/2\pi)\sqrt{H(t-t_0)}$ reflects the magnitude of the displacement
due to quantum fluctuations averaged over many Hubble volumes.  Even for the
case where
$|\phidot_q |\gg| \phidot_V|$ on average, there will be some regions where the
classical dynamics dominates the quantum. However the number of such
regions will be relatively small.

\subsection{Quadratic new inflation}
\label{quadnewinflssec}

We now consider a quadratic 
model of new inflation.
For quadratic new inflation with $V=V_0-m^2\phi^2/2$ the slow
roll conditions $|V''| \ll 9H^2$ and $|V' \mpl/V|\ll (48\pi)^{1/2}$
\cite{KT} give $m^2\ll9H^2$ and $\phi\ll (H^2/m^2)\mpl$.  We take
$\phi_e=V_0^\half/m$, as in Ref. \cite{LiddleLyth}.  From 
eqs. (\ref{phisqcoldwithphi0}), (\ref{phdot}) and (\ref{quantumcondn}), 
$\phi_Q=0.2 (H/m) H$.  Then from eq. (\ref{phisqcoldwithphi0}),
$N_Q=1.6 H^2/m^2$.  

For quadratic new inflation the WMAP bounds on $1-n_s$ imply GUT-scale
inflation \cite{LiddleLyth}.  Then
for $H=10^{-6}\mpl$, ${\cal N}=60$ and using eq. (\ref{N60})
\be
\phi_*=\phie e^{-20m^2/H^2}
\ee
The condition that $\phiQ<\phi_*$ then gives $m/H<0.8$. 
Setting ${\cal {R}}_k=H^2/(2\pi\phidot)_{*}=5\times10^{-5}$
and using $\phidot=-V'/3H$, one gets 
$m/H=0.03$ or $0.35$.  But $1-n_s=2m^2/(3H^2)$ \cite{LiddleLyth}
and WMAP observations imply that 
$0.025<1-n_s<0.049$ (at 68\% C.L.)\cite{jarosiketal}, thereby
allowing only $m/H=0.03$.  
Then $\phi_Q=7H$ and $N_Q=2000$.
Using eq. (\ref{phidotsqnew}) we can also note that 
$V(\phiQ)\approx V_0\gg\langle\phidot^2\rangle$ during the quantum phase, and so 
the
universe is 
potential energy dominated during and at the end of the fluctuation driven
phase.  
Thus the scenario with $\phi_Q<\phi_*$ is consistent.

\subsection{Other models of inflation}

We have also investigated the quantum phase of inflation for 
inflection point inflation, chaotic inflation
and natural inflation.
A quantum phase during inflection point inflation about the saddle point
$\bar\phi_0$ is mentioned
in Refs. \cite{al1,al2} 
(though the criterion for quantum evolution is
a bit different than ours)
and it is assumed that the field is out of the range for quantum evolution for the cosmologically
relevant phase of inflation.  We did not consider this scenario further.  The existence of the saddle
point requires a certain fine-tuned
relation between parameters in the lagrangian.  Refs. \cite{juan,EMS} consider
deviations from this relation.  In Ref. \cite{EMS} the deviation
is parametrised by a variable $\beta$.
For weak scale SUSY with $m\sim 100-1000\gev$, one has 
$\bar\phi_0\sim 10^{14-15}\gev$ 
and $V_0\sim 10^{32-34}\gev^4$,
and CMB observations require $\beta\sim O(10^{-10})$ \cite{EMS}.
(Also see Refs. \cite{juan,al2}.)  For such 
values 
the quantum phase condition in eq. (\ref{quantumcondn_a}) with $\phi_0$ replaced 
by $\bar\phi_0$
is valid only for
$|\phi-\bar\phi_0|\lsim 10^{-6}\gev$ whereas $H\sim 10^{-2,-3}\gev$.  Therefore 
one
can ignore the quantum phase for these models as the quantum phase range is much 
smaller than $H$.
Ref. \cite{EMS} also considers scenarios withput finetuning of $\beta$ but again
$|\phi-\bar\phi_0|\ll H$.

For quadratic and quartic chaotic inflation there is no quantum phase of 
inflation for $\phi<\phi_{QG}$ where $\phi_{QG}$ is the value of the $\phi$
for which quantum gravity effects become important $(V(\phi_{QG})=\mpl^4)$.
Natural inflation with a potential of the form
$V=\Lambda^4[1+\cos(\phi/f)]$ \cite{freese1,freese2} where
$\phi$ lies between 0 and $2\pi f$, and we take
$\phi<\pi f$, also does not have a quantum phase of inflation
(unless the field value is extremely small, $\phi/f<10^{-6}$).

\subsection{The quantum condition}

Another approach to identifying the quantum phase is to compare the evolution
of the inflaton due to quantum fluctuations and classical slow roll at an 
instant in time, over a time
interval $\Delta t=H^{-1}$ \cite{Linde,dimopetal}
i.e., to check if
\bea
\delta\phi_q = \frac{H}{2\pi} > \delta\phi_V&=& \dot\phi_V\, H^{-1}
\label{altquantumcondn}\\
&=& \frac{V'(\phi)}{3H^2}
\eea 
This approach checks if instantaneously 
the quantum jump overrides the classical evolution, while the condition
in eq. (\ref{phisqcoldwithphi0}) is a measure of whether quantum evolution 
dominates 
over classical evolution averaged over longer time durations. 
One notices that the condition in eq. (\ref{altquantumcondn}) 
is 
similar to
that for eternal inflation 
${\cal K}\equiv\delta\phi_q/[0.61\phidot_{V} H^{-1}] \gsim 1$ \cite{guth}.

Comparing the above approach and that in eqs. (\ref{phisqcoldwithphi0}) and
(\ref{phdot}), 
we see that the quantum evolution of the field in eq. (\ref{phisqcoldwithphi0}) 
goes
as $\phi_q(t) \sim \sqrt{{\cal N}(t)} = \sqrt{Ht}$, which means
${\dot \phi}_q \sim 1/\sqrt{{\cal N}} \sim 1/\sqrt{t}$.  
In
other words ${\dot \phi}_q$ decreases over time.
Thus if one is comparing this motion with the classical
motion and if the latter has approximately constant velocity,
as expected in a slow-roll regime, then it implies that at late time
eventually the classical motion always dominates.
Since the quantum kicks on the scalar field are a random
process, it might seem that there should be no dependence
on the past history of the quantum evolution that should
enter in comparing whether quantum or classical motion
dominates at a given time.  
In fact at any given instant
the RMS quantum kick in eq. (\ref{altquantumcondn})
is of order $H$, and so the RMS velocity
at any instant from quantum kicks is of order $H^2$.
So from this point of view, one might think the
relevant quantities to compare is whether this
RMS velocity from quantum evolution, which is independent
of past history, is the correct quantity to
compare against the classical velocity, as in eq. (\ref{altquantumcondn}).
However 
the history of the evolution 
should 
enter the comparison.
Even though at any given instant, the velocity from quantum
kicks is some approximately constant value, the direction is
random.  As such, with increaing time steps, there is an ever
increasing number if possible paths that the $\phi$ field
could have followed.  If one asks what
is the net motion from the quantum kicks after some
interval of time, on average it increases only
as $\sqrt{t}$. So a time derivative 
over an increasing
time interval actually is decreasing as $1/\sqrt{t}$.
On the other hand, the classical evolution of the field,
even if slow, is steadily always moving in the same direction.
As such, the longer one waits, the greater the chance that the $\phi$
field has arrived at increasing field values due to classical
rather than quantum evolution.

\section{$\phi_e>\phi_Q>\phi_*$}
\label{Qphase1}

We now consider the scenario where the
quantum fluctuations take the inflaton beyond $\phi_{*}$
and so our current horizon scale leaves the horizon when inflaton
evolution is dominated by quantum fluctuations. 
In this section we consider the scenario in which the fluctuations do not 
take the inflaton past $\phi_e$.
Since in the previous section we found that the quantum phase is
important only for small field inflation models  
here we investigate only new inflation and hilltop inflation models.

To obtain $\phi_{*}$ we 
break the evolution from $\phi_*$ to $\phi_Q$,
and from $\phi_Q$ to $\phie$.
Then
\bea
{\cal N}(\phi_{*}\rightarrow\phi_e)
&=&\int_{t_{*}}^{t_e} H dt\cr
&=& \int_{\phi_{*}}^{\phi_e} H \frac{d\phi}{\phidot}
\label{Nformula}\\
&=& \int_{\phi_{*}}^{\phi_Q} H \frac{d\phi}{\phidot}
+\int_{\phi_{Q}}^{\phi_e} H \frac{d\phi}{\phidot}
\label{Nformula1}\\
&=&
\frac{8\pi^2}{H^2}
\int_{\phi_{*}}^{\phi_Q} \phi \,
{d\phi}
+\frac{8\pi}{\mpl^2}\int_{\phi_{Q}}^{\phi_e} \frac{V(\phi)}{-V'(\phi)}{d\phi}
\label{N1N2}\\
&=&
\frac{4\pi^2}{H^2}(\phi_Q^2-\phi_*^2)
+\frac{8\pi}{\mpl^2}\int_{\phi_{Q}}^{\phi_e} \frac{V(\phi)}{-V'(\phi)}{d\phi}
\label{calN_1new}
\eea
where 
we have used eqs. (\ref{phisqcoldwithphi0}) and (\ref{phdot}) 
for $\phidot$ for $\phi<\phiQ$.

The comoving curvature perturbation is given by
\begin{equation}
{\cal{R}}=\psi_{com}=
\psi+{ H}\,\delta\tau
\end{equation}
where $\psi$ is the curvature potential, $\delta\tau=\delta\phi/\dot\phi$ is 
the time-displacement to go from a generic slicing with generic $\delta\phi$ to 
the comoving slicing with $\delta\phi_{com}=0$ \cite{riottolectures}.  
Evaluating
the rhs in the flat gauge in which $\psi=0$ gives
\begin{equation}
{\cal{R}}_k=
{H}\,\frac{\delta\phi_k}{\dot\phi}
\label{Rk_1}
\end{equation}
We first evaluate the rhs at horizon crossing.
In the flat gauge, $\delta\phi_k$ is given by $H/(2\pi)$.  $\dot\phi$ represents
the motion of the background condensate of long wavelength modes.  
Therefore
$\phidot$ is given by eq. (\ref{phidotsqnew}), 
Then
\begin{equation}
{\cal{R}}_k
= \frac{3H^3}{2\pi |V^{\prime\prime}(\phi_*)|\phi_*}
\label{Rkstar}
\end{equation}
where $\phi_*$ 
is given by
the value at horizon exit obtained from eq. (\ref{calN_1new}).
(The sign of ${\cal{R}}_k$ is suppressed as the relevant quantity is the
2-point function  $\sim |{\cal{R}}_k|^2$.)

While 
considering times after horizon exit
the background should only include those superhorizon modes that are of 
longer wavelength than the mode being investigated.  Therefore after horizon exit
both $\delta\phi_k$ and $\phidot$ in eq. (\ref{Rk_1})
evolve as in the standard classical roll picture and one gets constancy of
${\cal{R}}_k$ outside the horizon as usual.  (Quantum evolution of $\phi$ will 
continue as 
more and more modes cross the horizon and add to the condensate of 
fluctuations.  But the condensate value of $\phi_*$ relevant in 
eq. (\ref{Rkstar}) will not include modes that leave the horizon after
the mode being investigated.)

We point out an interesting fact that both $\delta\phi_k$ and $\dot\phi$
in ${\cal{R}}_k$ are quantum variables in the quantum phase of inflation.  
Therefore in the quantum phase 
we expect that there will be an increase in the cosmic variance because of the
stochastic nature of the evolution of $\phi$.

\subsection{Quartic new inflation}

For quartic new inflation, eq. (\ref{calN_1new}) implies
\be
\phi_*^2=\phi_Q^2-
\left[\frac{{\cal N}H^2}{4\pi^2}
-\frac{3H^2}{2(60\lambda)^\half}
+\frac{3}{8\pi^2\lambda^\half}\frac{H^4}{V_0^\half}
\right]\,.
\label{phistar}
\ee
The third term in the square brackets 
is larger than the second only for $H>2\mpl$.  Note that
limits such as $V_0^{1/4}<10^{16}\gev$ and $H<10^{13}\gev$ are 
derived for classical inflation and not valid in the quantum phase.
Nevertheless, $H>3\mpl$ implies $V(\phi)>\mpl^4$ which would put us in
the realm of quantum gravity.  
We take $H<\mpl$ and ignore the
third term in the square brackets above.

To check for consistency of this scenario, the quantity in square brackets
should be positive.  We also impose $\phi_*>H/(2\pi)$.  These conditions
give a lower bound and an upper bound on $\lambda$ respectively, namely, 
\begin{equation}
\frac{60}{{\cal N}^2}<\lambda<\frac{240}{ ({\cal N}+1)^2}
\,.
\end{equation}
(Imposing $\phi_Q<\phi_e$ only gives a weak bound $H<2\mpl$.)
For GUT-scale inflation with ${\cal N}\approx60$ this implies
\begin{equation}
0.02<\lambda<0.06 \,.
\label{lambdarange_newquartic}
\end{equation}
The cosmologically relevant scales leave inflation over 8 e-foldings.
For all these scales to leave during the quantum phase for GUT-scale
inflation the condition
is $0.02<\lambda$.
For inflation models with scales varying
from the electroweak scale to the Planck scale,
${\cal N}$ ranges from 30-65 and we get a lower bound of $(1-7)\times10^{-2}$ 
for $\lambda$ for all these models.
Clearly these values of $\lambda$ are interesting but one has to also check 
for constraints from the the power spectrum.

From eq. (\ref{Rkstar})
\be
{\cal{R}}_k=\frac{1}{2\pi}
\frac{H^3}{\lambda\,\phi_*^{3}}
\label{Rk_newquartic}
\ee
Using the expression for $\phi_*$ in eq. (\ref{phistar}) in 
eq. (\ref{Rk_newquartic}) we get
\begin{equation}
{\cal{R}}_k=
\frac{4\pi^2}{\lambda\,\Biggr[2\biggr(\frac{60}{\lambda}\biggr)^\half-{\cal{N}}
\Biggr]^\frac{3}{2}}
\label{Rk_newquartic1}
\end{equation}
For GUT-scale inflation with ${\cal N} =60$ we take $\lambda=4\times10^{-2}$
consistent with eq. (\ref{lambdarange_newquartic}) and get
\begin{equation}
{\cal{R}}_k \approx 13 \,.
\label{Rkvaluenewquartic}
\end{equation}
Such a large value of the curvature perturbation is ruled out by
observations.  Therefore this scenario, despite its mild constraint on
the inflaton coupling, namely, $\lambda\sim10^{-2}$, is not feasible.

It may be surprising that though in the quantum phase $\phidot$ is larger than 
the classical value, one gets such a large value of
${\cal{R}}_k$.  
However note that while $\dot\phi_q\gg\dot\phi_V$, 
$\sqrt{\langle\phidot^2\rangle}$ which enters the expression for ${\cal R}_k$
is $3\,\phidot_V$.  
Now the expression for ${\cal{R}}_k$ for classical
evolution, using eq. (\ref{Rk_1}) and $\dot\phi_V=-V'/(3H)$, is 
\be
{\cal{R}}_k=\frac{1}{2\pi}
\frac{3H^3}{\lambda\,\phi_*^{3}}\,.
\label{Rk_newquartic_cl}
\ee
Comparing this with eq. (\ref{Rk_newquartic}) one sees that the expressions are 
similar.
Yet, though
$\lambda$ in eq. (\ref{Rk_newquartic}) is large compared to $\lambda$ in 
eq. (\ref{Rk_newquartic_cl}),
${\cal{R}}_{k,q}\gg{\cal{R}}_{k,cl}$
because it happens that $\phi_*$ is much smaller than
in the standard classical evolution scenario.  
As shown earlier, the value of $\phi_*$ for the classical case 
for $\lambda=10^{-14}$ and ${\cal N}=60$ is $\phi_{*,cl}=10^6 H$.
However, for the
value of $\lambda=4\times10^{-2}$ and ${\cal N}=60$ one gets, 
from eq. (\ref{phistar}), 
$\phi_{*,q}=0.7H$.  

To understand why $\phi_*$ is much smaller for the quantum evolution scenario 
note that
i) $\phi_e\sim 1/\sqrt{\lambda}$ is much smaller in the quantum scenario than 
for the classical scenario as
$\lambda$ is large, and ii) a larger $\phidot_q$ in eq. (\ref{Nformula1}) 
requires a larger displacement in
$\phi$ to obtain the required number of e-foldings.  Both these effects push the 
inflaton
to a much smaller value of $\phi$ at the time of horizon exit 
for a given value of $\cal{N}$.

\subsection{Quadratic new inflation}

For quadratic new inflation eq. (\ref{calN_1new}) implies
\be
\phi_*^2=\phi_Q^2-
\left[\frac{{\cal N}H^2}{4\pi^2}
-\frac{3H^4}{4\pi^2m^2}\ln\biggr(\frac{\phi_e}{\phi_Q}\biggr)
\right]\,.
\label{phistar_quadnew}
\ee
Using expressions for $\phi_e$ and $\phi_Q$ from 
Sec. \ref{quadnewinflssec} we get 
\be
\phi_*^2=\phi_Q^2-
\left[\frac{{\cal N}H^2}{4\pi^2}
-\frac{3H^4}{4\pi^2m^2}\ln\biggr(\frac{1.7\mpl}{H}\biggr)
\right]\,.
\label{phistar_quadnew1}
\ee
The constraint that $\phi_Q>\phi_*$ implies
\be
\frac{m^2}{H^2}>\frac{3}{\cal N}
\ln\biggr(\frac{1.7\mpl}{H}\biggr)
\label{lowerboundonmnew}
\ee
while the requirement that $\phi_*>H/(2\pi)$ implies
\be
\frac{1.6}{{\cal N}+1} + \frac{3}{{\cal N}+1}
\ln\biggr(\frac{1.7\mpl}{H}\biggr)>\frac{m^2}{H^2}\,.
\label{upperboundonmnew}
\ee
Consistency of the above constraints implies
\be
H>1.7\mpl e^{-\frac{{\cal N}+1}{2}} 
\ee
For GUT-scale inflation with ${\cal N} = 60$ 
this requires $H>10^6\gev$.

To obtain the curvature perturbation we use eq. (\ref{Rkstar}) with $\phi_*$
given by eq. (\ref{phistar_quadnew}).
\be
{\cal R}_k = \frac{3}{2\pi}\frac{H^3}{m^2\,\phi_*}\,.
\label{Rkquadnew}
\ee
Now for $\phi_Q>\phi_*$ the curvature perturbation 
satisfies
\be
{\cal R}_k > \frac{3}{2\pi}\frac{H^3}{m^2\,\phi_Q} = 2 \frac{H}{m}
\label{Rkquadnewlimit}
\ee
For this to be consistent with observations would require $m/H>10^5$.
But such a large value of $m/H$ is in conflict with the upper bound
in eq. (\ref{upperboundonmnew}).  Therefore this scenario is not feasible.
(For electroweak scale inflation the lower bound on $H$ is too 
large ($3\times10^{12}\gev$) and so we do not consider it.)

\subsection{Hilltop inflation}

For the potential 
\begin{equation}
V=V_0 - \lambda \frac{\phi^p}{p \, M^{p-4}}\hspace{1cm} p>2 \,,
\label{potentialhilltop}
\end{equation}
$\cal N$
is obtained from
eq. (\ref{calN_1new}) as
\begin{equation}
{\cal N}
=
\frac{4\pi^2}{H^2}(\phiQ^2-\phistar^2)+
\frac{8\pi}{\mpl^2} \frac{V_0 M^{p-4}}{\lambda (p-2)}
\Biggr(
\frac{1}{{\phiQ}^{p-2}} - \frac{1}{\phi_e^{p-2}} 
\Biggr)
\end{equation}
and so 
\be
\phi_*^2=\phi_Q^2-
\left[\frac{{\cal N}H^2}{4\pi^2}
-
\frac{H^2}{4\pi^2}
\frac{8\pi}{\mpl^2} \frac{V_0 M^{p-4}}{\lambda (p-2)}
\Biggr(\frac{1}{\phiQ^{p-2}} - \frac{1}{\phi_e^{p-2}} \Biggr)
\right]\,.
\label{phistarhilltop1}
\ee
Using $H=[(8\pi/3)V_0/\mpl^2]^{1/2}$
\be
\frac{1}{\phiQ^{p-2}} - \frac{1}{\phi_e^{p-2}}
=
\frac{1}{
\left[\frac{H^2}{\mpl^2}\frac{V_0 M^{p-4}}{\lambda}\right]^\frac{p-2}{p}}
-
\frac{1}{
\left[\frac{V_0 M^{p-4}}{\lambda}\right]^\frac{p-2}{p}}
\ee
Considering $H^2\ll\mpl^2$, we ignore $1/\phi_e^{p-2}$.
For $p>4$ and $H\ll M$ (the potential in eq. (\ref{potentialhilltop}) is an 
effective
potential for $\phi<M$, 
and we only consider $\phi>H/(2\pi)$) 
one gets from eq. (\ref{Rkstar}) (using Mathematica)  
\be
{\cal{R}}_k
=0.5 \frac{1}{\lambda^\frac{1}{p}}\left(\frac{M}{H}\right)^\frac{p-4}{p}
\ee
which is greater than 1.

\section{$\phi_Q>\phi_e$}
\label{Qphase2}

We now consider the scenario where evolution in the inflationary phase is
entirely dominated by quantum fluctuations, i.e., $\phi_Q>\phi_e$.  Then
\be
{\cal N}(\phi_{*}\rightarrow\phi_e)
=\frac{4\pi^2}{H^2}(\phi_e^2-\phi_*^2) \,,
\ee
and so
\be
\phi_*=\left[\phie^2 - \frac{{\cal N}}{4\pi^2} H^2\right]^\half\,.
\label{phi60forphiQphie}
\ee

\subsection{Quartic new inflation}

For quartic new inflation, as discussed earlier, $\phi_e\approx(V_0/\lambda)^{1/4}$.
The condition
$\phi_Q>\phi_e$ implies that 
\be
H>1.8\, 
\mpl\,.
\ee 
(This is close to the limit $H<3\mpl$ for classical gravity to be valid.)
From eq. (\ref{phi60forphiQphie}), $\phi_e^2>{\cal N} H^2/(4\pi^2)$.
(Imposing $\phi_*>H/(2\pi)$ gives a similar bound with 
${\cal N}\rightarrow {\cal N} +1$.)
For Planck scale inflation we take ${\cal N}=65$.
This then implies 
\be
H< \frac{
0.02 \,\pi^2}{\sqrt\lambda}\,\mpl\,.
\ee
Combining the above two bounds one finds that for inflation to be in
the quantum phase during its entire duration requires a large 
Hubble parameter during inflation and only a reasonable upper bound on
$\lambda$, i.e.,
\be
H>1.8\,\mpl \hspace{1cm}
{\rm and}
\hspace{1cm}
\lambda<1\times10^{-2}\,.
\label{quarticlimits}
\ee
The comoving curvature perturbation is given by 
eq. (\ref{Rk_newquartic}) with $\phi_*$ as in eq. (\ref{phi60forphiQphie}).
Once again we obtain a large value of ${\cal R}_k\,(>17)$ which rules out this
scenario.

\subsection{Quadratic new inflation}

The condition
that $\phi_Q>\phi_e$ implies
\be
H>1.7\mpl
\label{quadnew1}
\ee
and the condition that $\phi_e^2>{\cal N} H^2/(4\pi)$ in eq. (\ref{phi60forphiQphie})
implies that
\be
\frac{3}{2{\cal N}}\mpl^2 > m^2\,.
\label{quadnew2}
\ee
For Planck scale inflation ${\cal N} = 65$, and $0.15 \mpl>m$.
The curvature perturbation in eq. (\ref{Rkquadnew}) satisfies
\be
{\cal R}_k > \frac{3}{2\pi}\frac{H^3}{m^2\,\phi_e}\,.
\label{Rkquadnewineq}
\ee
Using eqs. (\ref{quadnew1}) and (\ref{quadnew2}) we get ${\cal R}_k > 26$ which is in conflict
with observations.

\section{Other Scalar Fields}
\label{Otherscalars}

\subsection{Curvaton}

We now consider scenarios where a field other than the inflaton 
evolves due to its quantum fluctuations during inflation.
In the curvaton scenario, quantum fluctuations of a field $\sigma$ with
a flat potential is responsible for the density perturbations in the Universe
\cite{lythwands,LUW}.  (Also see 
Refs.\cite{mollerach,lindemukh,moroitakahashi1,moroitakahashi2}.)
The curvature perturbation in these scenarios is 
given by 
\begin{eqnarray}
\zeta&=&\frac{4\rho_r \zeta_r + 3\rho_\sigma \zeta_\sigma}
{4\rho_r + 3\rho_\sigma}\\
&\approx&r\zeta_\sigma
\end{eqnarray}
where the subscript $r$ refers to radiation, and the variable $r$ is the ratio of
the curvaton energy density $\rho_\sigma$ to the 
total energy density
$\rho$
just before the curvaton decays.  One presumes that 
$\zeta_r$ is negligible and it is ignored.

The curvature spectrum due to the curvaton field is obtained from the expression 
for
$\zeta_\sigma$ at the time when the curvaton starts to oscillate 
after inflation when $H$ falls to $m_\sigma$ at $t_{osc}$.
\be
{\zeta}_{\sigma,k}=\frac{2}{3}\frac{\delta\sigma_k}{\sigma}|_{osc}\,,
\label{stdcurvR}
\ee
Since $\delta\sigma_k$ is constant outside the horizon, we take 
$\delta\sigma_k(t_{osc})$ to be
$H/(2\pi)$.  We take $\sigma(t_{osc})$ 
to be the value $\sigma_e$ at the end of 
inflation at $t_e$.
$\sigma_e=\sigma_{cl}(t_e)\pm \delta\sigma_q(t_e)$.  If 
$\sigma_e$ is determined by quantum fluctuations then
\be
{\zeta}_{\sigma}=\frac{2}{3}\frac{\delta\sigma_k}{\delta\sigma(t_e)}\,.
\label{curv-with-q-new}
\ee
For a curvaton mass $m_\sigma$,
$\delta\sigma(t_e)$ is either $[H/(2\pi)]\,N_e^\half$ or $3H^4/(8\pi^2 m_
\sigma^2)$,
depending on whether the condensate of curvaton fluctuations does not or 
does
attain the asymptotic value by $t_e$.  ($N_e$ is the total number of e-foldings of
inflation.) If it does not, then
%
\be
{\zeta}_{\sigma} = \frac{2}{3} \frac{1}{N_e^\half}\,.
\ee
Now $\zeta_\sigma=\zeta/r$ where $\zeta=5\times10^{-5}$,
and $10^{-2}<r<1$.  (The lower limit on $r$
comes from the relation between $r$ and the non-gaussianity parameter
$f_{NL}$, namely $f_{NL}=5/(4r)$ \cite{LUW},
and the upper limit on $f_{NL}$ of $O(100)$ \cite{komatsuetal2011}.)  This then implies
\begin{equation}
2\times10^4<N_e<2\times10^8\,.
\end{equation}
Therefore curvaton models where quantum fluctuations determine $\sigma_e$ and 
where $\delta\sigma$ has not 
reached the asymptotic value by $t_e$ require more e-foldings of inflation 
than standard models
of inflation.  For quartic new inflation models the upper limit on $N_e$ is of the same
order as the duration of the quantum phase.

The condition that the curvaton has not reached its asymptotic limit (for a quadratic
curvaton potential) is \cite{lindefluc}
\begin{equation}
\frac{2 m_\sigma^2}{3H^2}N_e\ll1\,.
\end{equation}
Combining this with the lower bound on $N_e$ above gives
\begin{equation}
\frac{m_\sigma^2}{H^2}\ll 8\times10^{-5}\,.
\end{equation}
Now the scalar power spectrum spectral index in the curvaton scenario is
\cite{LUW} 
\begin{equation}
n_s=1-2\epsilon_H + 2\eta_{\sigma\sigma}
\end{equation}
where  
$\epsilon_H = -\dot H/H^2$ and $\eta_{\sigma\sigma}=\mpl^2/(8\pi)
\frac{\partial^2 V/\partial\sigma^2}{V}$.  For a quadratic 
curvaton potential, $\eta_{\sigma\sigma} = m_\sigma^2/(3H^2)$.  Then for 
$n_s=0.963$ \cite{komatsuetal2011} 
\begin{equation}
\frac{m_\sigma^2}{3H^2}\approx\epsilon_H-0.02\,.
\label{msigmaH}
\end{equation}
Therefore the above upper bound on $m_\sigma^2/H^2$ further implies that 
in quadratic curvaton models where quantum fluctuations determine 
$\sigma_e$
and 
$\delta\sigma$ has not 
reached the asymptotic value by 
$t_e$,
$\epsilon_H= 0.02$.
\footnote{This was also pointed out to us by K. Dimopoulos.}
Small field inflation
models 
such as
new inflation, small field natural
inflation and some hybrid inflation
models 
with a concave downward potential
would not satisfy this 
criterion \cite{BL,alabidilyth}.

If $N_e\gg (3H^2)/(2m_\sigma^2)$ 
one uses the asymptotic value for the curvaton 
fluctuations in Eq. (\ref{curv-with-q-new}).  Then using 
the observed value of $\zeta$ and
the bounds on $r$ one finds 
\begin{equation}
9\times10^{-5}<
\frac{m_\sigma}{H}<
9\times10^{-3}\,.
\label{curv-msigmaH}\end{equation}  
This is in conflict with the bound $m_\sigma/H\lsim 10^{-4}$ obtained in 
Sec. IIIB of Ref. \cite{postma} from CMB constraints.  (We believe there is
an error in the derivation of the bound in Ref. \cite{postma}.)
Once again, Eq. (\ref{msigmaH}) implies that $\epsilon_H= 0.02$ which is 
incompatible 
with small field inflation models.  The lower bound on $H/m_\sigma$ from 
Eq. (\ref{curv-msigmaH}) and the lower bound on $N_e$ above also implies that 
inflation must last much longer than $10^4$ e-foldings.

If 
$\sigma_e$ is determined by $\sigma_{cl}(t_e)$ rather than by quantum fluctuations
then $\delta\sigma_q(t_e)\ll \sigma_e$ and
\be
{\zeta}_{\sigma}\ll\frac{2}{3}\frac{\delta\sigma_k}{\delta\sigma(t_e)}\,.
\label{curv-with-q-ll}
\ee
and one can only conclude that $N_e<2\times10^8$ and $m_\sigma/H>9\times10^{-5}$ 
respectively 
in the scenarios where the curvaton fluctuations do not and do
attain their asymptotic value.

As an aside, we comment on certain existing results in the literature
which may be relevant for the curvaton scenario.
The asymptotic value of the quantum fluctuations for the curvaton
\be
\langle\sigma^2\rangle = \frac{3 H^4}{8\pi^2 m_\sigma^2}\,.
\label{asympcurva}
\ee
may be used in eq. (\ref{stdcurvR})
as in Refs. \cite{postma}. 
However it may be shown (via eq. (4.11) of Ref. \cite{FKL}, eq. (13) of
Ref. \cite{FMSVV} and eq. (14) of Ref. \cite{deharoelizalde})
that on including the time variation
of $H$ during quadratic chaotic inflation 
the asymptotic value of the fluctuations during inflation of any light
scalar field
goes as 
\be
\langle\sigma^2\rangle = \frac{3 H^4}{16\pi^2 m^2}\,,
\label{asympcurv1a}
\ee
where $m$ is the inflaton mass.
If one uses eq. (\ref{asympcurv1a}) then 
one gets 
\be
{\zeta}_{\sigma}=\left(\frac{2}{3}\right)^{\frac{3}{2}} \frac{m}{H}
\approx 1.6\times10^{-2}
\ee
which would need $r\approx3\times10^{-3}$ to agree with the observed density
perturbations.  
Such
a small value of $r$ will imply $f_{NL} \approx 400$ 
which 
is in conflict with  
bounds on $f_{NL}$ \cite{komatsuetal2011}.
In models with non-quadratic terms in the curvaton potential 
$f_{NL}$ can be small even when
$r$ is 
small \cite{enqnurmi,sasakietal,enqtaka,ENTT} because the coefficient of
the term proportional
to $1/r$ in $f_{NL}$ in the presence of non-quadratic terms 
can be very small.
These models are then not ruled out
(though Ref. \cite{ENTT} indicates that very small values of $r$ are
not favoured).

\subsection{Quintessence}

In quintessence models the dark energy is associated with a slowly rolling 
background field $Q$.  Since the potential must be flat so that potential energy
dominates the kinetic energy (and the pressure is negative) we verify whether
quantum evolution can dominate over the slow classical evolution.
Consider the potential 
\be
V=V_0\left(\frac{\mpl}{Q}\right)^p
\label{quintpot}
\ee
The condition in eq. (\ref{quantumcondn}) implies that quantum evolution 
dominates for 
\be
\frac{H^2}{4\pi\sqrt{{\cal N}}} > \frac{p}{8\pi}\left(\frac{\mpl}{Q}\right) 
H \mpl
\ee
(We have presumed that the de Sitter result for the fluctuations is valid in
our current epoch of acceleration.)
For $Q\sim\mpl$ \cite{mliddle} and $H\ll\mpl$ the above inequality is not
satisfied.  
Note that any perturbations generated in the current accelerating phase leave the
horizon and can not be detected.

In the model of quintessential inflation \cite{peeblesvilenkin} the quintessence
field plays the role of the inflaton at early times.  The potential is as in 
eq. (\ref{quintpot}) with $p=4$.  The value of $Q$ today is approximately its 
value at the beginning of the radiation dominated era and is about $8\mpl$.
Once again, for $H\ll\mpl$ quantum evolution does not dominate during the 
current accelerated phase.

In Ref. \cite{ringevaletal} the dark energy today is 
associated with a frozen condensate of fluctuations of a field $\varphi$
generated during inflation. 
The field, which has a quadratic potential, is almost massless during inflation 
and
evolves due to quantum fluctatuations, similar to the
evolution of the inflaton in the quantum phase that has been studied above.  
As the Universe evolves after inflation
the superhorizon fluctations remain frozen as a condensate at a value
$\varphi_c=[3H_I^4/(8\pi^2 m^2)]^{1/2}$
and 
dominate the energy density of the Universe today ($H_I$ is the Hubble parameter 
during inflation).   
The condensate
behaves like a slowly rolling quintessence field today with the equation of 
state of
dark energy.  Inflation in this model occurs at a low scale of $5 \tev$.
The curvature perturbation generated during inflation due to quintessence field
fluctuations is given by eq. (\ref{Rkstar})
which implies 
\begin{eqnarray}
{\cal R}_k 
&=&\frac{3H^2}{m^2}\frac{\delta\varphi}{\varphi}\\
&=&\sqrt{6} H_I/m\gg1\,.
\end{eqnarray}

CMB fluctuations are influenced by the curvature perturbation at decoupling. 
The contribution of the dark energy to the curvature perturbation at decoupling 
will be 
${\cal R}_k(t_{dec})*f $ where
\bea
f
&=&\frac{(\rho+p)_\varphi}{(\rho+p)}|_{dec}\\
&=&\frac{\dot\varphi^2}{\rho_{dm}}|_{dec}\\
&=&\frac{m^2 H_I^4}{9\pi H_{dec}^4\mpl^2}
\eea
and we have used $\dot\varphi=-m^2\varphi/(3H)$ and approximated $\varphi$ by
the value $\varphi_c$ above (as also in Ref. \cite{ringevaletal}). 
To obtain the value of the curvature 
perturbation at decoupling
one assumes that $\delta\varphi/\varphi$ remains constant during slow roll 
because both
$\delta\varphi$ and $\varphi$ have the same equation of motion (as in
curvaton models).  Once the field starts oscillating when
$H\sim m$, then ${\cal R}_k\sim \delta\rho/\rho|_\varphi \sim
\delta\varphi/\varphi \sim$ constant. Therefore
${\cal R}_k(t_{dec})=\sqrt{6} (H_I/m) (m/H_I)^2$.
Then
\bea
f{\cal R}_k(t_{dec}) 
&=&\frac{\sqrt{6}m^3 H_I^3}{9\pi H_{dec}^4\mpl^2}\\
&\ll& 5\times10^{-5}\,,
\label{quintcondR}
\eea
where $m\lsim H_0$, the Hubble parameter today, 
as indicated in this model.
Thus the curvature perturbation at decoupling due to the quintessence
condensate will not give rise to a large net curvature perturbation.
Therefore the curvature perturbation due to the quintessence field condensate
in this model is not in conflict
with CMB observations.

\section{Warm Inflation}
\label{WI}

In warm inflation
\cite{Berera:1995ie},
dissipative effects are important during inflation
so that radiation production occurs concurrently with inflationary
expansion.
The basic equation for describing the evolution
of an inflaton field that dissipates energy
is of a Langevin form \cite{Berera:1995wh,Berera:1999ws}
\begin{equation}
{\ddot \phi} + [3H + \Upsilon] {\dot \phi}
- \frac{1}{a^2(t)} \nabla^2 \phi
- \frac{\partial V}{\partial \phi} = \zeta.
\label{wieom}
\end{equation}
In this equation, $\Upsilon {\dot \phi}$ is
a dissipative term and $\zeta$ is a fluctuating force.
Both are effective terms, arising
due to the interaction of the inflaton
with other fields.  In general these two terms are
related through a fluctuation-dissipation theorem,
which would depend on the statistical
state of the system and the microscopic dynamics.
Although the statistical state can be quite general,
all studies so far have focused on the thermal
state and we will restrict our consideration here to
that also.  Thus
the evolution of the inflaton field has to
be calculated in a thermal background.

The above dynamics need not be restricted just to the period when there
is a potential driven inflation period. The above dynamics could also
occur previous to such a period. This point leads to a new type of
inflation phase in which while inflation occurs,
the inflaton rather than being governed by the potential $V$
is instead governed by the thermal background which produces 
large fluctuations in the inflaton field.
This is similar to the quantum fluctuation driven inflation 
discussed in the previous sections,
except now the fluctuations are thermal rather than quantum.

To examine the initial period during this thermal fluctuation dominated
inflation phase,
we first need to evaluate $\phi_T$ which is the value of $\phi$ when
evolution due to the fluctuations is no longer dominant.
Following the same procedure as done for the cold inflation
case, we need to evaluate the equivalent of
eq. (8.3.12)
of
Ref.\cite{Linde}
and
eq. (3.11)
of Ref. \cite{BST}
to obtain $\phidot$ due
to fluctuations and due to the potential and ascertain till when the
former dominates.
Fluctuations in warm inflation are obtained from a Langevin equation
derived using a real-time formalism of thermal field
theory \cite{Berera:1999ws,gleiserramos}.

In treating warm inflation, one caveat is important.
In general the inflaton dynamics is a non-equilibrium problem.
Whether the case of cold or warm inflation, certain assumptions are
already being made about this dynamics when one writes down
the evolution equation.  In cold inflation, the basic assumption
is the inflaton is evolving at effectively zero temperature
and interactions with other fields are negligible.
In the case of warm inflation, one is assuming
there is a thermal state and the inflaton interaction with
other fields is significant.  In this case there will
be a point in time, $t_d$, when these conditions are realized.
Previous to that time, in principle one would need
to calculate the full quantum field dynamics and
determine the statistical state and its evolution.
This is beyond the scope of this paper.  Here what we will assume is
either dissipation effects are important and evolve
as eq. (\ref{wieom}) or else they are not important and to
a good approximation the evolution is the same as the
cold inflation case.  If $t_d$ is sufficiently early,
then the entire fluctuation dominated era can be calculated
based on the evolution eq. (\ref{wieom}).  However if $t_d$
occurs fairly close to the onset of the slow-roll
period, then
this can allow for possibilities which combine
both quantum and thermal fluctuation regimes before
potential driven inflation.

There are two dissipative regimes that must be considered
depending on whether in eq. (\ref{wieom})  $\Upsilon \leq 3H$, which
is called the weak dissipative regime,
or $\Upsilon > 3H$, which is called the strong dissipative
regime.  For all these cases, the Langevin
equation for the modes of the inflaton field can
be solved, as done in Ref. \cite{HMB}.
Our interest here is in the long wavelength modes,
for which Ref. \cite{HMB} finds the solution coincides
with the homogeneous solution, as if the effect of
the noise force had no effect.  Thus calculation of
$\langle \delta\phi^2 \rangle$ in both these cases is found
to have the same general form as for cold inflation
and, following eqs. (7.3.10-7.3.12) of
Ref.\cite{Linde},
\begin{equation}
\langle \delta\phi^2 \rangle
=\frac{1}{(2\pi)^3}\int_H^{aH} {d^3k}   |\phi_k|^2
=\int_{H}^{aH} \frac{dk}{k} P_\phi(k) \ ,
\label{phi2pk}
\end{equation}
where only the inflaton power spectrum
$P_\phi(k)\equiv k^3 |\phi_k|^2/(2\pi^2)$
is different for the various cases.
The limits
of the integral admit only those modes that have left the
horizon during inflation.  Below we consider a quartic new inflation potential.

\subsection{Weak dissipation}

In the scenario in which $\Upsilon<3H$ the slow roll of the inflaton
is because of the Hubble damping term in the equation of motion,
so in particular the inflaton mass $m < 3H$ must hold
in order that a slow-roll regime is eventually achieved.
The analysis for this case is very similar to the cold
inflation case, except the expression for the inflaton
fluctuation is different.  In this regime the
inflaton fluctuation at freeze-out gives 
$P_\phi(k) = 
(3\pi/4)^{1/2}
HT$
\cite{Berera:1995wh,Berera:1999ws,HMB}
and
\begin{equation}
\langle\delta\phi^2\rangle
=
\left(\frac{3 \pi}{4}\right)^\frac{1}{2}
\, H T \times N(t)\,.
\label{phithw}
\end{equation}
(This may be compared with eq. (\ref{phisqcold}), 
$\langle\delta\phi^2\rangle=(H/2\pi)^2 \times N(t)$.)
Then, for $\phi_0\approx0$,
\be
\phi = 
\left(\frac{3 \pi}{4}\right)^\frac{1}{4}
( H T)^\frac{1}{2} 
\sqrt {H(t-t_0)} \ ,
\label{phith_wk}
\ee
and the evolution rate of the inflaton in the thermal fluctuation
driven phase is
\be
\phidot_{\rm{th}} =
\frac{1}{2}
\left(\frac{3 \pi}{4}\right)^\frac{1}{4}
H^\frac{3}{2} T^\frac{1}{2}  
\frac{1}{\sqrt {H(t-t_0)}} \,.
\label{phidot_thw}\ee
Thermal fluctuations will dominate the evolution of $\phi$ as long as
\be
\frac{1}{2}
\left(\frac{3 \pi}{4}\right)^\frac{1}{4}
H^\frac{3}{2} T^\frac{1}{2}  
\frac{1}{\sqrt {H(t-t_0)}}
\gg
-\frac{V'}{3H}
=\frac{\lambda\phi^3}{3H} \,.
\ee
Defining $\phi_T$ as the largest value for which thermal fluctuations dominate,
$\phi_T \approx H^\frac{3}{4} T^\frac{1}{4}/\lambda^\frac{1}{4}$.

The temperature during the thermal phase must be determined.
Any residual radiation from initial conditions will rapidly
redshift away during any inflation epoch, and so in
order for a thermal fluctuation dominated phase of
inflation to  exist, there must be a source of radiation production.
Although in general this is a problem of nonequilibrium quantum
field theory, following our statements at the start of this section,
we will assume the radiation is produced
by the background component of the scalar field, $\phi$, which
is controlled by 
eq. (\ref{wieom}) without the noise force, for which the radiation produced is
\begin{equation}
\rho_r = \frac{\Upsilon {\dot \phi}^2}{4H}.
\label{rhoreq}
\end{equation}
In the weak dissipative regime, ${\dot \phi} = -V'/(3H)$,
and for the
quartic
new inflation potential of Sec. \ref{CI}
this gives
\begin{equation}
\rho_r = \frac{\lambda ^2 \Upsilon \phi^6}{36 H^3}.
\label{rhorwni}
\end{equation}
Since the radiation energy density increases with $\phi$,
an estimate of the largest radiation energy
density produced will be for $\phi \sim \phi_T$.
Equating the expression eq. (\ref{rhorwni}) with
$\rho_r \approx g_* T^4$, the temperature can be
expressed in terms of the other quantities, and
we find $T \sim 0.2 \lambda^{1/5} \Upsilon^{2/5} H^{3/5}/g_*^{2/5} \ll H$.
The latter inequality follows since $\lambda \ll 1$
and in the weak dissipative
regime $\Upsilon < H$. 
Thus we conclude that in the weak dissipative case for the new
inflation quartic potential, there is never a thermal fluctuation driven
regime during inflation.  The dynamics before potential driven inflation will
be the same as for the quantum fluctuation dominated
phase in cold inflation.

\subsection{Strong dissipation}

In the strong dissipation case
$P_\phi(k) = \sqrt{\pi/4} (\Upsilon H)^\half T$ \cite{Berera:1999ws,HMB}
(see eq. (34) of Ref. \cite{HMB}).
Results will now be calculated
during the fluctuation era for the
quartic
new inflation model. 
Using the above expression for $P_\phi(k)$ in eq. (\ref{phi2pk}),
$\langle\delta\phi^2\rangle$ in the fluctuation dominated regime can be obtained as
\begin{eqnarray}
\langle\delta
\phi^2\rangle
&=&
\left(\frac{\pi}{4}\right)^\half
(\Upsilon H)^{1/2} T \times N(t) \,.
\end{eqnarray}
Then, for $\phi\approx0$,
\be
\phi =
\left(\frac{\pi}{4}\right)^\frac{1}{4}
(\Upsilon H)^\frac{1}{4} T^\frac{1}{2} \sqrt {H(t-t_0)} \ ,
\label{phith}
\ee
and so
\be
\phidot_{\rm{th}} \approx (\Upsilon H)^{\frac{1}{4}} T^\frac{1}{2} H \frac{1}{2\sqrt {H(t-t_0)}} .
\label{phidot_th}\ee
The
thermal fluctuations will dominate the $\phi$ evolution from
the potential as long as
\be
(\Upsilon H)^\frac{1}{4} T^\half H \frac{1}{2\sqrt {H(t-t_0)}} \gg
\frac{\lambda\phi^3}{\Upsilon} .
\label{flucdom}\ee
If fluctuations dominate evolution till $\phi_T$ then
eq. (\ref{flucdom}) implies
that
\begin{equation}
\phi_T=
(\Upsilon H)^{3/8} T^{1/4}
/\lambda^{1/4}
\,.
\label{phiT}
\end{equation}

The temperature during the thermal fluctuation dominated phase
must now be determined using eq. (\ref{rhoreq}), where in
the strong dissipative regime ${\dot \phi} = -V'/\Upsilon$.
To estimate the maximum the radiation energy density will
be during the thermal fluctuation inflation regime,
eq. (\ref{phiT}) is used from which we find,
\begin{equation}
T \sim \frac{\lambda^{1/5}}{g_*^{2/5}} (\Upsilon H)^{1/2} .
\label{Tsdwi}
\end{equation}
In order for this regime to be thermal dominated requires
$T > H$, which implies the condition
\begin{equation}
\Upsilon \stackrel{>}{\sim} \frac{g_*^{4/5}}{\lambda^{2/5}} H .
\end{equation}
In this regime, from eqs. (\ref{phith}) and (\ref{phiT})
\be
N_T=
\frac{(\Upsilon H)^\frac{1}{4}}{(\lambda T )^\half}\,.
\ee

The curvature power spectrum for strong dissipation 
is given by \cite{Berera:1999ws,HMB}
\bea
P_R^\half&=&
\frac{H\Upsilon}{|V'|} P_\phi^\half\\
&=&
2 \pi^\frac{1}{4}
\frac{(\lambda T)^\half}
{(H\Upsilon)^\frac{1}{4}}N_k^\frac{3}{2} ,
\eea
with $N_k\approx H\Upsilon/(2\lambda\, \phi_k^2)$.  (The expression for
$N_k$ is that for the weak dissipative regime with $3H$ replaced
by $\Upsilon$, as can be surmised from eq. (\ref{Nformula}).)
Setting $N_k=60$ and $P_R^\half=10^{-5}$ gives 
$N_T=(H \Upsilon)^\frac{1}{4}/(T\lambda)^\half= 10^8$.
One can verify that
$V(\phi_T)
\approx V_0- (\Upsilon/H)^\frac{3}{2} T H^3/4 \approx
V_0 - \lambda^{1/5} (\Upsilon H)^2/(4 g_*^{2/5})$.
In order that $V(\phi_T) \approx V_0$ and 
so the fluctuations do not take the inflaton to the
bottom of the potential, it requires
the condition $10 V_0/(\lambda^{7/20} M_{Pl})^4 \ll 1$,
which holds for $V_0^{1/4} \stackrel{<}{\sim} 10^{14} {\rm GeV}$.
This is two orders of magnitude
less than earlier constraints based on CMBR density fluctuation 
measurements \cite{BasteroGil:2009ec}.

In this regime, we also must confirm that the kinetic energy
of the field fluctuations does not dominate the
potential during the fluctuation driven epoch.
The kinetic energy of the field fluctuations is obtained from
\be
\langle
{\phidot}^2\rangle = \int_{k_F}^{ak_F} \frac{dk}{k} 
|\dot {\delta \phi}(k)|^2 \, ,
\label{sdpd}
\ee
where 
$\delta\phi(k)
=(k^3/2\pi^2)^\half\,\phi_k$ and
$k_F = (\Upsilon H)^{1/2}$ is the freeze-out scale in
the strong dissipative regime \cite{Berera:1999ws}.
Using eq. (38) of Ref. \cite{HMB}
for $\dot{\delta\phi}(k)$, 
\be
\dot{\delta\phi}(k)
=-\left(\frac{\pi}{4}\right)^\frac{1}{4}
\frac{k^2}{\Upsilon a^2} (\Upsilon H)^\frac{1}{4} T^\half\, ,
\ee
we get from eq. (\ref{sdpd}),
\be
\langle
{\phidot}^2\rangle \approx
\Upsilon^{1/2} H^{5/2} T ,
\ee
which is much smaller than $V_0 = 0.1 H^2 \mpl^2$
since $T$ and $\Upsilon$ are much smaller than $\mpl$.
Also from eq. (\ref{Tsdwi}) it follows that $\rho_r$
is much less than $V_0$.

This result has demonstrated a new thermal fluctuation inflation
phase.  There can be other variations to the above scenario.
In the strong dissipative regime slow-roll motion
only requires the condition on the inflaton mass
$m < \Upsilon$, thus in general the inflaton mass
can be much bigger than the Hubble parameter.
This feature, combined with the presence of
a radiation component and dissipative dynamics can lead
to other possible dynamics previous to the potential
driven inflation phase.
One example is
if the dissipation dynamics is not initially active,
previous to time $t_d$, the inflaton field
now is massive and evolves like the cold
inflation case, in that only a $3 H {\dot \phi}$ term
damps its evolution.    In such a case
for a  massive inflaton field the estimate
for $\langle \delta\phi^2 \rangle$ differs from
that in eq. (\ref{phisqcold}) and  rather it
is eq. (7.3.13) of Ref. \cite{Linde}.

\section{Conclusion}
\label{Concl}

In this article we have 
investigated
the evolution of the inflaton due to
quantum fluctuations and studied its possible consequences.
If the field travels far on its potential due to quantum fluctuations
or acquires a 
large kinetic energy 
then standard inflation where the field rolls
due to the slope of its potential will not subsequently occur.  
This will dramatically alter the inflationary scenario and significantly
affect the density spectrum.
For a given potential a priori one can not predict 
the impact of quantum fluctuations on the 
evolution of the inflaton.
For example, for GUT-scale inflation
the quantum phase lasts for $10^8$ e-foldings 
for a Coleman-Weinberg potential (approximated by a quartic
potential).  
In contrast, for quadratic new inflation the 
quantum phase lasts for 2000 e-foldings.  
For chaotic inflation, inflection point inflation and for natural inflation, the quantum phase
is negligible and classical rolling is
important from the beginning of inflation.
One might have expected quantum fluctuations to be relevant for all
small field inflation models but it is not so for small field natural
inflation.

For new inflation models, if one
assumes that our current horizon scale left the horizon during classical slow
roll, then the 
earlier
quantum phase 
ends with the inflaton far from
the minimum of its potential and with sub-dominant kinetic energy.  This
allows for the standard classical rolling inflationary phase to follow.
If cosmologically relevant scales leave
the horizon during the 
quantum
phase, which is subsequently followed by a classical phase, then for quartic new 
inflation
this requires that the coupling $\lambda$ is 
greater than $10^{-2}$.
We have derived the expression for the curvature perturbation which is valid
for the quantum phase.   
We get a large value for the curvature
perturbation for modes that leave during the quantum phase.  This then rules
out this scenario.  
We also consider a scenario where the quantum phase lasts for the
entire inflationary epoch 
which is also ruled out because of the large curvature perturbations.
Our conclusions are similar for quadratic new inflation.

We have also studied curvaton and quintessence models where quantum 
evolution can be relevant.  
{For curvaton models, if the 
curvaton evolves during inflation
due to quantum fluctuations which determine the curvaton field value
at the end of inflation, 
we find that the number of e-foldings
of inflation must be orders of magnitude more than the usual minimum value 
required to
solve the horizon problem.  We also find that the slow roll parameter 
$\epsilon_H$ is
determined and has a value that is incompatible with small field inflation 
models. 
We further point out that newer
results for asymptotic values of scalar fields during inflation can lead to
large non-gaussianity in conflict with observations.  Quantum fluctations
in quintessence models do not lead to any inconsistencies with observations
though in the condensate dark energy model the curvature perturbation associated
with the condensate field is large during inflation.

We have studied
quartic new inflation in the context of warm inflation (weak
dissipation and strong dissipation regimes).
We find that as in cold inflation about $10^8$ e-foldings
of inflation occur before inflaton evolution is driven by the slope
of the potential.  
However in the weak dissipative regime, the fluctuation driven
phase is due to quantum fluctuations while in the 
strong dissipative regime it is due to thermal fluctuations.
Quantum fluctuations of the inflaton in a thermal background
are larger than in vacuum, 
and the condition that the inflaton is not driven to the minimum
of its potential 
by fluctuations in the strong dissipative regime requires that
the scale of inflation must be less than $10^{14}\gev$.  
In both dissipative regimes the kinetic energy
of the inflaton at the end of the fluctuation driven phase is much
less than the potential energy, thereby allowing for the standard
warm inflation scenario with a slowly rolling inflaton field driven by
the potential to commence after the fluctuation driven phase is over.
This confirms the robustness of the warm inflation scenario.

\begin{acknowledgements}
A.B. would like to thank the Physical Research Laboratory, Ahmedabad, India
for support and hospitality while researching this article.
We would like to thank the organisers of the Xth Workshop on High Energy
Physics Phenomenology (WHEPP-X) at the Institute of Mathematical Sciences,
Chennai, India during which discussions on the above work were 
initiated.  We would also like to thank Mar Bastero-Gil,  Karim Malik 
and Kostas Dimopoulos for very useful discussions.  We
also acknowledge very useful comments from an anonymous referee.

\end{acknowledgements}

\end{document}